\documentclass[12pt,a4paper]{article}
\usepackage[dvips]{graphicx}
\begin{document}
\title{Nonlinear Model of non-Debye Relaxation}
\author{Boris A. Zon\\ Voronezh State University,\\ University Sq.1, 394006 Voronezh, Russia
}
\maketitle
\begin{abstract}
We present a simple nonlinear relaxation equation which contains
the Debye equation as a particular case. The suggested relaxation
equation results in power-law decay of fluctuations. This equation
contains a parameter defining the frequency dependence of the
dielectric permittivity similarly to the well-known one-parameter
phenomenological equations of Cole-Cole, Davidson-Cole and
Kohlrausch-Williams-Watts. Unlike these models, the obtained
dielectric permittivity (i) obeys to the Kramers-Kronig relation;
(ii) has proper behaviour at large frequency; (iii) its imaginary
part, conductivity, shows a power-law frequency dependence
$\sigma\sim\omega^n$ where $n<1$ corresponds to empirical
Jonscher's universal relaxation law while $n>1$ is also observed
in several experiments. The nonlinear equation proposed may be
useful in various fields of relaxation theory.
\end{abstract}

The frequency dependence of the dielectric permittivity
$\varepsilon(\omega)$ plays a crucial role in numerous aspects 
of condensed matter science. Among the great number of
publications, we can mention the textbooks \cite{Book,Jons-96} and
review articles \cite{Jons-99,RMP}. Classical theory of polar
dielectric relaxation, which is connected with the
$\varepsilon(\omega)$, is based on the linear Debye equation
\begin{eqnarray}\label{tauD}
\frac{dW}{dt}=-\frac{1}{\tau}(W-W_0).
\end{eqnarray}

Here $W$ is the angular distribution of dipoles, $W_0$ is its
equilibrium value and $\tau$ is a relaxation time. The solution of
the Eq. (\ref{tauD}) describes an exponential relaxation, which
results in the Debye dielectric permittivity
\begin{eqnarray}\label{epsD}
\varepsilon(\omega)=\varepsilon_\infty+\epsilon_D(\omega),\quad
\epsilon_D(\omega)=\frac{\varepsilon_s-\varepsilon_\infty}
{1+\imath\omega\tau},
\end{eqnarray}
where $\varepsilon_\infty$ is the dielectric permittivity at
$\omega\gg1/\tau$ and $\varepsilon_s\equiv \varepsilon(0)$.
However, the dependence (\ref{epsD}) occurs quite seldom in
experiments \cite{Book,Jons-96,Jons-99,RMP}. Explanation of a
non-Debye dispersion $\varepsilon(\omega)$ is carried out in
frames of the models which can be tentatively divided into three
directions. (i) Models which assume the relaxation time $\tau$ to
be a stochastic variable. Taking different distributions for
$\tau$ one can obtain from the equation (\ref{epsD}) different
$\varepsilon(\omega)$ dependencies, including those matching the
experimentally observed curves (see, for instance,
Refs.~\cite{Weron,Ven}). Apparently this theoretical approach to
explain the experimental data is not faultless. (ii)
Electrical network model used for the microstructural media
\cite{PRL-04}. (iii) Models which consider influence of manybody
effects on the relaxation \cite{Book,RMP,DH1,DH2}. The model
suggested here is close to the latter type.

The kinetic equation for spatial distribution of dipoles includes
a collision integral which model is right hand side of Eq.
(\ref{tauD}) (see, for instance, Ref. \cite{Gross}). In general
kinetic theory this integral is a nonlinear functional of the
distribution. It is therefore natural to take the right hand side
of the relaxation equation as a nonlinear function. The simplest
nonlinear generalization of the Debye equation (\ref{tauD}) is the
following relaxation equation:
\begin{eqnarray}\label{tau1}
\frac{dW}{dt}=-\frac{(W-W_0)^{q+1}}{\tau}.
\end{eqnarray}

The parameter $q$ in the equation (\ref{tau1}) defines the
relaxation law, shape of absorption line and dispersion dependence
for the real part of $\varepsilon$. In this regard the parameter
$q$ is analogous to the parameters of Cole-Cole, Davidson-Cole,
Kohlrausch-Williams-Watts phenomenological curves \cite{Ven}. In
the present case the fitting parameter is introduced at an earlier
stage, i.~e., into a differential relaxation equation. This leads
to a number of principal differences comparing to the
aforementioned phenomenological curves.

A nonlinear relaxation equation
\begin{eqnarray}\nonumber
\frac{dW}{dt}=-\frac{W^{q+1}-W_0^{q+1}}{\tau},
\end{eqnarray}
which is similar to the Eq. (\ref{tau1}), also includes the only
shape parameter $q$, though it leads to more sophisticated
mathematical formulae.

\begin{figure*}
\includegraphics[scale=1.0]{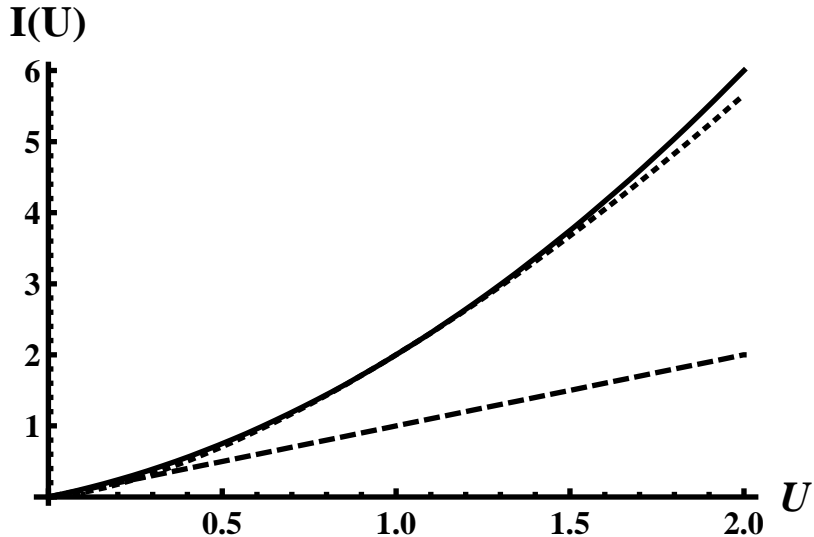}
\caption{Hypothetical  view of the collision integral $I(U)=U+U^2$
(solid curve) and approximation of it in the Debye linear model
$I(U)\simeq U$ (dashed curve) and in the proposed nonlinear model
$I(U)\simeq 2U^{1.5}$ (points). Multiplier 2 in the latter
equation defines the difference between the relaxation times in
the linear and nonlinear models.} \label{fig.1}
\end{figure*}

Obviously, for noninteger $q$ the right hand side of the
Eq.~(\ref{tau1}) displays nonanalytic (in the sense of
Cauchy--Reimann conditions) behaviour which cannot follow from
kinetic theory. For small deviations from the equilibrium,
\begin{eqnarray}\label{U}
U=W-W_0,
\end{eqnarray}
the right hand side of the relaxation equation should have Debye
form~(\ref{tauD}). However, this is a rather difficult question
whether $U$ can be considered as small enough. We assume that the
$U$ values relevant to the experiment are not small so that the
relaxation equation can be approximated by the simple
formula~(\ref{tau1}) (see Fig.~\ref{fig.1}).

From the Eq. (\ref{tau1}) one easily obtains the relaxation
function
\begin{eqnarray}\label{U1}
U(t)=\left(qt/\tau+C\right)^{-1/q},
\end{eqnarray}
where $q$ is positive number to satisfy the condition $W\to W_0$
at $t\to\infty$. The constant $C=U_0^{-q}$ ensures the initial
condition $U(0)=U_0$. Then
\begin{eqnarray}\label{U3}
U(t)=U_0(1+qtU_0^q/\tau)^{-1/q}.
\end{eqnarray}

Function (\ref{U3}) coincides with the Debye relaxation function
$U(t)=U_0\exp(-t/\tau)$ for $q\to +~0$.


The general relation between the electric field $E(t)$ and the
electric displacement field $D(t)$ will be used. The principle of
causality gives for the isotropic media \cite{LL}
\begin{eqnarray}\label{D}
D(t)=\varepsilon_\infty E(t)+ \int_{-\infty}^t \kappa(t-\theta)
E(\theta)d\theta,
\end{eqnarray}
where $\kappa$ may differ from the relaxation function $U(t)$ only
by a constant positive multiplier \cite{F}
\begin{eqnarray}\label{kappa}
\kappa(t)=A(1+t/\tau^*)^{-1/q},
\end{eqnarray}
and an effective relaxation time
\begin{eqnarray}\label{tau*}
\tau^*=\frac{\tau}{qU_0^q}
\end{eqnarray}
is introduced.

Multiplier $A$ can be calculated for $q<1$ from the equation
(\ref{D}) with the dc field $E$. Taking into account the equation
$D=\varepsilon_s E$ for this case,
\begin{eqnarray}\label{A}
A=\frac{1-q}{q\tau^*} (\varepsilon_s-\varepsilon_\infty), \quad
q<1.
\end{eqnarray}

For $q>1$ the field $E$ cannot be assumed as a dc since integral
in the equation (\ref{D}) diverges. Multiplier $A$ cannot be
calculated in this case. This problem is discussed below.

To determine an ac permittivity $\varepsilon(\omega)$ let us
consider a periodic field $E(t)=E_0e^{-\imath\omega t}$ in the Eq.
(\ref{D}). Then
\begin{eqnarray}\label{eps}
\varepsilon(\omega)=\varepsilon_\infty+ \epsilon(\omega),
\end{eqnarray}
\begin{eqnarray}\nonumber
\epsilon=\epsilon_1 +\imath \epsilon_2 = A\int_0^\infty
(1+t/\tau^*)^{-1/q} e^{\imath\omega t}dt \\\nonumber =\frac{\imath
A}{\omega}(-\imath \omega\tau^*)^{1/q} e^{-\imath \omega\tau^*}
\Gamma\left(1-1/q, -\imath \omega\tau^*\right),
\end{eqnarray}
where $\Gamma(\cdot,\cdot)$ is an incomplete $\Gamma$-function
\cite{GR}. In a more common form  one can write $\epsilon(\omega)$
as
\begin{eqnarray}\label{eps1}
\epsilon=\epsilon_1 -\imath \epsilon_2 = -\frac{\imath
A}{\omega}(\imath \omega\tau^*)^{1/q} e^{\imath \omega\tau^*}
\Gamma\left(1-1/q, \imath \omega\tau^*\right).
\end{eqnarray}

It is easy to see the dependence $\epsilon$ on $\omega$ is
determined by effective relaxation time $\tau^*$. The actual
relaxation time, $\tau$, which enters the Eqs. (\ref{U1}),
(\ref{U3}), can be measured by the methods of time domain
spectroscopy \cite{Cole,Feldman}. Since $\epsilon(\omega)$ is
Fourier-Laplace transform of a smooth function it obeys to the
Kramers-Kronig relation \cite{F}. Function $\epsilon(\omega)$ has
an integrable singularity at $\omega=0$ when $q>1$ and this
singularity is an inaccuracy of the model. In fact, it is well
known that behaviour of a function in frequency domain near
$\omega=0$ is determined by the function in time domain at
$t\to\infty$. However, for very large $t$ the function $W$ is
close to the function $W_0$ and equation (\ref{tau1}) becomes
incorrect. So the equation (\ref{eps1}) cannot be used with $q>1$
in the narrow frequency region near $\omega=0$.

Imaginary part of the permittivity, $\epsilon_2$, is connected with
the conductivity of a matter, $\sigma$, by the well-known
equation
\begin{eqnarray}\label{sigma}
\sigma(\omega)=\frac{\omega}{4\pi}\epsilon_2(\omega)
\end{eqnarray}
and $\sigma(\omega)$ has no singularity at $\omega=0$ for any $q$.
Its behaviour at small $\omega$ is the following
\begin{eqnarray}\label{sigma2}
\displaystyle
\sigma(\omega) \simeq\left\{
\begin{array}{ll}\frac{A}{4\pi}\Gamma(1-1/q) \cos(\pi/2q)
\,(\omega\tau^*)^{1/q},\, q>1/2\\
\frac{A(\omega\tau^*)^2} {4\pi(1-2q)},\, q<1/2\end{array}\right.
\end{eqnarray}

It is interesting to note that the dependence
$\sigma\sim\omega^2$, which follows from the Eq. (\ref{sigma2})
for $q<1/2$, coincides with its dependence in Debye theory when
$q=0$. The dependence $\sigma\sim\omega^n$ with $n<1$ has been
observed in many experiments (Jonscher's universal relaxation law
\cite{Jons-96, Jons-99,Jons-77,Jons-81}). Such a dependence
follows from the equation (\ref{sigma2}) for $q>1$. Dependence
$\sigma\sim\omega^n$ with $n>1$ which follows from the equation
(\ref{sigma2}) for $1/2<q<1$ has been also observed in some types
of materials, e.~g., in glassy
0.3(xLi$_2$O.(1-x)Li$_2$O)0.7B$_2$O$_3$ \cite{mat1}, in mixed
compounds of (NH$_4)_3$H(SO$_4$)$_{1.42}$(SeO$_4)_{0.58}$
\cite{mat2} and K$_3$H(SeO$_4)_2$ single crystals \cite{mat3}.
Sometimes at very low frequencies, $\omega<\omega_0$, in both
($n<1$ and $n>1$) cases the frequency dependence of conductivity
is weak, plateau-like. It does not contradict the proposed
model since in the low-frequency region the conductivity may be
connected with other processes, magnitude of which becomes equal
to (\ref{sigma2}) when $\omega\simeq\omega_0$~\cite{Papa}.

For $\omega\to\infty$
\begin{eqnarray}\label{as}
\epsilon_1\simeq\frac{A}{\omega^2\tau^*}, \quad \epsilon_2
\simeq\frac{A}{\omega},
\end{eqnarray}
that coincides with the asymptotic behavior of $\epsilon_D$
(\ref{epsD}). We remind that the dependence
$\epsilon_1\sim\omega^{-2}$ is a general feature of a dielectric
permittivity at large frequencies \cite{LL}. Nevertheless, for
many materials the decrease of $\epsilon_1$ with the increasing
frequency is slower than $\omega^{-2}$. It may be connected with
the influence of faster relaxation processes which induce extra
high-frequency maxima on the dielectric losses curve.

\begin{figure*}
\includegraphics[scale=1.0]{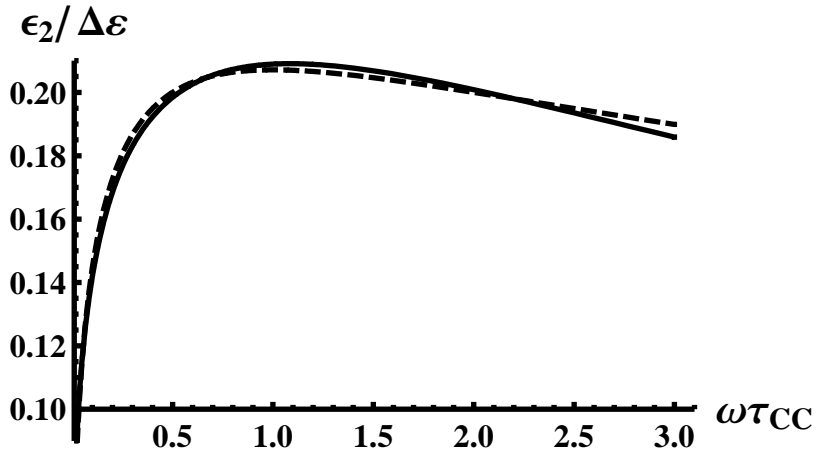}
\caption{Imaginary part of the dielectric permittivity,
$\epsilon_2$, calculated according to the equation (\ref{eps1})
(solid curve) and to the Cole-Cole equation (\ref{CC1}) (dashed
curve). All parameters are given in the text.} \label{fig.2}
\end{figure*}


Clearly, the suggested model requires a detailed experimental
verification. Here we demonstrate that it may be adjusted with the
well-known empirical dependencies, e.~g., with the Cole-Cole
equation
\begin{eqnarray}\label{CC1}
\epsilon_{CC}=\frac{\Delta\varepsilon}
{1+(\imath\omega\tau_{CC})^\alpha}, \quad
\Delta\varepsilon\equiv\varepsilon_s-\varepsilon_\infty.
\end{eqnarray}
Figure \ref{fig.2} shows the frequency dependence of $\epsilon_2$
for $\alpha=1/2$. The parameters of nonlinear relaxation are
chosen as $q=0.732$ and $\tau^*=4.66\tau_{CC}$. Such a large
difference of $\tau^*$ and $\tau_{CC}$ implies a shift of
dielectric losses curve maximum to the low-frequency region.

In conclusion, the nonlinear relaxation equation is proposed. The
equation leads to a power decay of the fluctuations. Frequency
dependencies of the dielectric permittivity and dielectric losses
are in agreement with the well-known empirical dependencies.
Derived relaxation equation may be useful to describe different
relaxation processes. If a relaxation of the dimensional values is
under consideration, $U$ in (\ref{U}) should be equal to
$W/W_0-1$.

\end{document}